\documentclass[twocolumn,aps,pra,amsmath,amssymb,showpacs]{revtex4}
\usepackage[latin9]{inputenc}
\usepackage{float}
\usepackage{amsmath}
\usepackage{amssymb}
\usepackage{graphicx}
\usepackage{esint}

\makeatletter

\newcommand*\LyXThinSpace{\,\hspace{0pt}}

\@ifundefined{textcolor}{}
{%
 \definecolor{BLACK}{gray}{0}
 \definecolor{WHITE}{gray}{1}
 \definecolor{RED}{rgb}{1,0,0}
 \definecolor{GREEN}{rgb}{0,1,0}
 \definecolor{BLUE}{rgb}{0,0,1}
 \definecolor{CYAN}{cmyk}{1,0,0,0}
 \definecolor{MAGENTA}{cmyk}{0,1,0,0}
 \definecolor{YELLOW}{cmyk}{0,0,1,0}
}

\usepackage{epsfig}\usepackage{amsfonts}

\makeatother

\begin{document}

\title{Optomechanical devices based on traveling-wave microresonators}

\author{Yan-Lei Zhang, $^{1,2}$ }

\author{Chun-Hua Dong, $^{1,2}$ }

\author{Chang-Ling Zou, $^{1,2,3}$ }
\email{clzou321@ustc.edu.cn}

\author{Xu-Bo Zou, $^{1,2}$ }
\email{xbz@ustc.edu.cn}

\author{Ying-Dan Wang, $^{4,5}$}

\author{Guang-Can Guo $^{1,2}$ }

\affiliation{$^{1}$ Key Laboratory of Quantum Information, University of Science
and Technology of China, Hefei, 230026, People's Republic of China; }

\affiliation{$^{2}$ Synergetic Innovation Center of Quantum Information \& Quantum
Physics, University of Science and Technology of China, Hefei, Anhui
230026, China}

\affiliation{$^{3}$ Department of Applied Physics, Yale University, New Haven,
CT 06511, USA}

\affiliation{$^{4}$ CAS Key Laboratory of Theoretical Physics, Institute of Theoretical
Physics, Chinese Academy of Sciences, P.O. Box 2735, Beijing 100190,
China}

\affiliation{$^{5}$ School of Physical Sciences, University of Chinese Academy
of Sciences, No.19A Yuquan Road, Beijing 100049, China}

\date{\today}
\begin{abstract}
We theoretically study the unique applications of optomechanics based
on traveling-wave microresonators, where the optomechanical coupling
of degenerate modes can be enhanced selectively by optically pumping
in different directions. We show that the unique features of degenerate
optical modes can be applied to the entangled photon generation of
clockwise and counter-clockwise optical modes, and the nonclassicality
of entangled photon pair is discussed. The coherent coupling between
the clockwise and counter-clockwise optical mods and two acoustic
modes is also studied, in which the relative phase of the optomechanical
couplings plays a key role in the optical non-reciprocity. The parity-time
symmetry of acoustic modes can be observed in the slightly deformed
microresonator with the interaction of forward and backward stimulated
Brillouin Scattering in the triple-resonance system. In addition,
the degenerate modes are in the decoherence-free subspace, which is
robust against environmental noises. Based on parameters realized
in recent experiments, these optomechanical devices should be readily
achievable.
\end{abstract}

\pacs{42.50.Wk, 42.50.Ex, 07.10.Cm, 11.30.Er}
\maketitle

\section{Introduction}

Optomechanical devices \cite{Aspelmeter2014,Kippenberg2008,Aspelmeyer2012,Thourhout2010,Hong2013}
have attracted considerable attentions in integrated photonic circuits
\cite{Tang1,Eichenfield2009,Fan2016}, memory \cite{Tang2,Cole2011}
and high precision measurements \cite{Anetsberger2010,Weber2016,Ockeloen-Korppi2016}
for practical applications. There are also excellent test bed for
investigating the quantum behaviors at macroscopic level \cite{Purdy2013}
and are attractive for fundamental studies on physics, such as the
gravitational wave detection \cite{Aspelmeter2014,Aasi2013,Belenchia2016},
quantum-to-classical transitions \cite{Ludwig2008,Brennecke2008},
and quantum information processing \cite{Khalili2010,Stannigel2012}.
In past few years, many remarkable progresses have been achieved both
in theory and experiment, such as ground state cooling of the mechanical
resonator \cite{Teufel2011}, optomechanically induced transparency
\cite{Weis2010,Safavi-Naeini2011}, optomechanical entanglement \cite{Palomaki2013}
and squeezing \cite{Wollman2015}, and optical frequency conversion
\cite{Dong2012,Andrews2014}. 

Recently, more and more studies on multimode optomechanics (MOM) \cite{Aspelmeter2014}
have been carried out, for more functional devices and applications.
Because they provide more degrees of freedom, people can realize new
phenomena that are not possible in single mode optomechanical systems,
such as the optical non-reciprocity \cite{Xu2015,Meltelmann2015,Peano2016},
parity-time symmetry \cite{Peng2014}, chiral symmetry breaking \cite{Wurl2016},
and topological energy transfer \cite{HXu2016}. Different from the
triple-resonance enhanced optomechanical interaction \cite{phononlaser,Zhu},
the MOM in this paper is referring to more than one optical signal
mode or mechanical mode. Many schemes \cite{Stannigel2012,Buchmann2015,Kipf2014,Chesi2015}
have been proposed to realize the MOM \cite{Deng2016} based on the
coupled cavities or hybrid optomechanical systems. However, in practical
physical systems, it is challenging to fabricate multiple optical
or mechanical resonators that have the frequencies matching each other.
In additional to the frequency mismatching problem, the MOM also requires
complex design for the efficient coupling between separated resonators. 

In this paper, we propose to use the traveling-wave microresonators
to explore the MOM, which allows many potential new phenomena and
applications. The whispering-gallery microcavities \cite{Dong2012,Park2009,Balram2014,Baker2014,Grudinin2009,Bahl2011,Bahl2012}
is one of the candidate systems, which supports degenerate clockwise
(CW) and counter-clockwise (CCW) traveling-wave modes, which have
already been used in optomechanically induced non-reciprocity \cite{Shen2016,Ruesink2016,Kim2015,Dong2015}.
By using the degenerate oppositely propagating optical modes, we demonstrate
the entangled photon generation and the controllable optical non-reciprocity.
The CW and CCW acoustic waves also enable the study of the non-Hermitian
dynamics of phonons, and the parity-time symmetry of acoustic modes
is proposed in the slightly deformed traveling-wave microresonator.
The traveling-wave microresonator provides an excellent platform for
studying the optomechanics with multiple degrees of freedom, and the
schemes proposed in this work are feasible for experiments. The integration
of all modes in single solid state microresonator also holds the great
advantage that the noise of degenerate CW and CCW traveling-wave modes
could cancel each other, where the mode pair actually forms a decoherence-free
subspace, which makes optomechanical devices robust against environmental
noises and holds great potential for applications in quantum information
processing.

The paper is organized as follows. In Sec. II, we show the CW and
CCW traveling-wave modes in traveling-wave microresonators, and develop
the Hamiltonian description of the MOM. Based on this model, we propose
three unique applications in this system: the entangled photon generation
of CW and CCW modes in Sec. III, the non-reciprocal conversion between
CW and CCW optical modes in Sec. IV, and the phononic parity-time
symmetry in Sec. V. Based on parameters realized with existing technology
in recent experiments, these optomechanical devices should be readily
achievable. We give a discussion about the potential advantages of
the traveling-wave optomechanical system in In Sec. VI, and then summarize
in Sec. VII.

\section{The system}

\begin{figure}
\includegraphics[width=8cm]{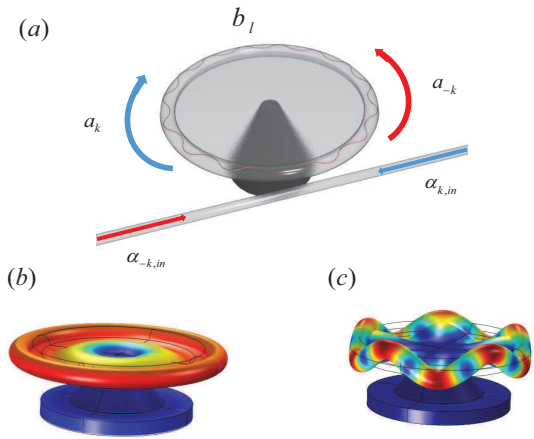}

\caption{(Color online) (a) Schematic of the the travelling-wave multimode
optomechanical system. Here $a_{k}$ and $a_{-k}$ are the annihilation
operators of the clockwise (CW) and counter-clockwise (CCW) traveling-wave
modes, and $b_{l}$ is the annihilation operator of the mechanical
mode. (b) The breath mechanical mode $l=0$ for dispersive optomechanics.
(c) The Brillouin mechanical mode $l=5$ for triple-resonant optomechanics.}
\end{figure}

In this work, we consider an MOM system as shown in Fig.$\,$1 (a):
an integrated non-magnetic dielectric microresonator served as both
optical and mechanical resonator. The microresonator possesses the
rotational symmetry, leading to the degeneracy of the optical or mechanical
modes that are propagating in the CW and the CCW directions. In general,
the Hamiltonian description of the system can be written as 
\begin{equation}
H=H_{0}+H_{i}+H_{d},
\end{equation}
in which ($\hbar=1$)
\begin{equation}
H_{0}=\sum_{k=-N}^{N}\omega_{c,k}a_{k}^{\dagger}a_{k}+\sum_{l=-M}^{M}\omega_{m,l}b_{l}^{\dagger}b_{l}
\end{equation}
is the system Hamiltonian, which describes the optical and mechanical
modes. Due to the rotational symmetry, both optical and mechanical
mode fields can be represented by the traveling field $\overrightarrow{E}_{k}\left(r,z\right)e^{-ik\phi}$
and $\overrightarrow{u}_{l}\left(r,z\right)e^{-il\phi}$, with $r,z,\phi$
representing the cylindrical coordinator, $k$ and $l$ are the azimuthal
quantum numbers. Therefore, we use $a_{k}$ and $b_{l}$ as the annihilation
operators of the $k$-th optical and $l$-th mechanical modes, and
the $\omega_{c,k}$ and $\omega_{m,l}$ are the corresponding mode
frequencies. As mentioned above, we have $\omega_{c,k}=\omega_{c,-k}$
and $\omega_{m,l}=\omega_{m,-l}$ due to the CW and CCW symmetry.

The mechanical motion of the dielectric microresonator can modify
the optical mode frequency, through the geometry effect and the photo-elastic
effect \cite{Balram2014,Baker2014}, which leads to the optomechanical
interaction Hamiltonian as 
\begin{equation}
H_{I}=\sum_{j,k,l}g_{j,k,l}\left(a_{j}+a_{j}^{\dagger}\right)\left(a_{k}+a_{k}^{\dagger}\right)\left(b_{l}+b_{l}^{\dagger}\right).
\end{equation}
The coupling strength $g_{j,k,l}$ is determined by the overlap of
the optical electric field and the mechanical displacements or strain
\cite{Balram2014,Baker2014,Tomes2011}. Since the integral $\int d\phi e^{-i\left(j+k+l\right)\phi}=\delta\left(j+k+l\right)$,
the optomechanical interaction in this traveling-wave microresonator
should satisfy the selection rule that $j\pm k\pm l=0$. We can separate
the interaction Hamiltonians into two categories: 
\begin{enumerate}
\item Dispersive optomechanics as shown in Fig.$\,$1 (b). 
\begin{equation}
H_{I,1}=\sum_{k}g_{k,k,0}a_{k}^{\dagger}a_{k}\left(b_{0}+b_{0}^{\dagger}\right).
\end{equation}
Here, $g_{k,k,0}$ is the single-photon coupling rate, which corresponds
to the optical cavity frequency shift per phonon excitation. This
dispersive optomechanics indicates the mechanical motion shifts the
optical mode frequency. Due to the selection rule, only the mechanical
mode with $l=0$, i.e. the breath-type modes \cite{Weis2010,Shen2016}
can give rise to such interaction. 
\item Triple-resonant optomechanics as shown in Fig.$\,$1 (c). 
\begin{equation}
H_{I,2}=\sum_{j,k}g_{j,k,j-k}\left(a_{j}^{\dagger}a_{k}b_{j-k}+a_{j}a_{k}^{\dagger}b_{j-k}^{\dagger}\right).
\end{equation}
It describes a Brillouin scattering in this triple-resonant system
\cite{Dong2015}, where a photon in mode $j$ split into a photon
in mode $k$ and a phonon in mode $j-k$. Here, $g_{j,k,l}$ is the
Brillouin scattering coupling strength, which is non-zero only when
the energy $\left(\omega_{m}=\omega_{c,j}-\omega_{c,k}\right)$ and
momentum $\left(l=j-k\right)$ conservation are satisfied.
\end{enumerate}
The term $H_{d}$ describes the driving of the optical modes and also
the optical signal input to the system. With the weak interaction
coupling, we can use strong optical driving fields to enhance optomechanical
coupling strength \cite{Dong2012,Dong2015}, and the optomechanical
system evolves by optical signal input.

\section{Degenerate photon pair generation}

For dispersive optomechanics, we can choose the two degenerate CCW
($-k$) and CW ($k$) optical modes that are coupled to the breath
mechanical mode, thus the interaction Hamiltonian is
\begin{equation}
H=g_{k,k,0}\left(a_{k}^{\dagger}a_{k}+a_{-k}^{\dagger}a_{-k}\right)\left(b_{0}+b_{0}^{\dagger}\right).
\end{equation}
The system is a general bosonic three-mode system, as shown in the
Fig.$\,$2(a). Such model can also be realized in a ring type resonator
coupling to a local mechanical oscillator \cite{Aspelmeter2014}.
Since both $a_{k}$ and $a_{-k}$ can coherently couple with the mechanical
mode, it is possible to realize the interaction between $a_{k}$ and
$a_{-k}$ mediated by the mechanical mode. It has been demonstrated
that the red-detuned laser pump on both CW and CCW direction can induce
the effective coupling between the CW and CCW signal photons. If we
pump one direction with blue detuned laser, there will generate a
photon-phonon pair (anti-Stokes process), and the red detuned laser
on the other direction will convert the phonon to photon (Stokes process),
thus generate entangled photons propagate in opposite directions.

To study the entangling interaction between CW and CCW signal photons,
we assume two driving fields of the CW and CCW optical modes with
frequency detuned from the cavity as $\Delta_{d,k'}=\omega_{d,k'}-\omega_{c,k'}$,
with $k'=\pm k$, where $\omega_{d,k'}$ is the drive laser frequency.
Let the detuning $-\Delta_{d,k}=\Delta_{d,-k}=\omega_{m}$, and apply
the standard linearization treatment, the effective Hamiltonian can
be written as
\begin{align}
H_{lin}= & -\Delta_{d,\mathrm{k}}a_{\mathrm{k}}^{\dagger}a_{k}-\Delta_{d,\mathrm{-k}}a_{\mathrm{-k}}^{\dagger}a_{\mathrm{-k}}+\omega_{m}b_{0}^{\dagger}b_{0}\nonumber \\
 & +G_{\mathrm{k}}\left(a_{\mathrm{k}}^{\dagger}b_{0}+a_{\mathrm{k}}b_{0}^{\dagger}\right)+G_{\mathrm{-k}}\left(a_{-k}^{\dagger}b_{0}^{\dagger}+a_{\mathrm{-k}}b_{0}\right).
\end{align}
Here, the coupling strength $G_{\mathrm{k}\mathrm{\left(-k\right)}}=g_{k,k,0}\alpha_{\mathrm{k}\mathrm{\left(-k\right)}}$
is enhanced by the driving field amplitude $\alpha_{\mathrm{k}\mathrm{\left(-k\right)}}$.
, The rotating wave approximation has been applied above with the
assumption $\omega_{m}\gg G_{k\mathrm{\left(-k\right)}},~\kappa,~\gamma_{m}$,
where $\kappa$ and $\gamma_{m}$ are the cavity and mechanical damping
rates, respectively. 

For studying the entanglement of the photon pair, a weak signal is
sent into CW optical mode as 
\begin{equation}
H_{s}=i\sqrt{\kappa_{in}}\epsilon_{s,k}\left(a_{\mathrm{k}}^{\dagger}e^{-i\left(\omega_{s}-\omega_{d,\mathrm{k}}\right)t}-H.c.\right),
\end{equation}
where $\kappa_{in}$ is the external coupling to the cavity. The dynamics
of the density matrix $(\rho)$ of the system is governed by the Master
equation \cite{Scully1997}, which reads
\begin{align}
\frac{d\rho}{dt}= & -i\left[H_{lin}+H_{s},\rho\right]+\kappa\left[\mathcal{L}\left(a_{\mathrm{k}}\right)+\mathcal{L}\left(a_{-k}\right)\right]\nonumber \\
 & +\left(n_{th}+1\right)\gamma_{m}\mathcal{L}\left(b_{0}\right)+n_{th}\gamma_{m}\mathcal{L}\left(b_{0}^{\dagger}\right),
\end{align}
where $\mathcal{L}\left(A\right)=A\rho A^{\dagger}-\left(A^{\dagger}A\rho+\rho A^{\dagger}A\right)/2$
is the Lindblad superoperator for any operator $A$, and $n_{th}$
is the phonon thermal excitation. In our calculation, we choose the
damping rate of optical mode $\kappa/2\pi=15\,\mathrm{MHz}$, the
damping rate of mechanical mode $\gamma_{m}/2\pi=22\,\mathrm{kHz}$,
and $\kappa_{in}=\kappa/2$ from the experimental parameters \cite{Shen2016}.

\begin{figure}[H]
\includegraphics[width=9cm]{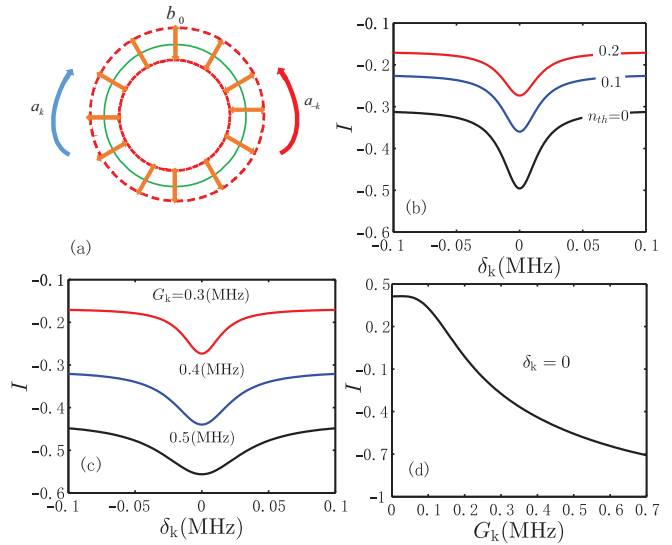}

\caption{(Color online) (a) Schematic diagram of entangled photon generation.
(b) The parameter $I$ as a function of the detuning $\delta_{\mathrm{k}}$
for $G_{\mathrm{k}}/2\pi=0.3$ MHz and $n_{th}=0,~0.1,~0.2$. (c)
The parameter $I$ as a function of the detuning $\delta_{k}$ for
$n_{th}=0.2$ and $G_{k}/2\pi=0.3~0.4,~0.5$ MHz. (d) The parameter
$I$ as a function of the effective coupling $G_{\mathrm{k}}$ for
the detuning $\delta_{\mathrm{k}}=0$ and $n_{th}=0.2$. Other parameters
are $G_{\mathrm{-k}}/2\pi=0.1$ MHz and $\epsilon_{s,k}=0.1$ MHz.}
\end{figure}

To characterize the nonclassicality of the two-mode field, we introduce
the parameter \cite{Agarwal1988} 
\begin{equation}
I=\frac{\sqrt{\left\langle a_{\mathrm{k}}^{\dagger2}a_{\mathrm{k}}^{2}\right\rangle \left\langle a_{\mathrm{-k}}^{\dagger2}a_{\mathrm{-k}}^{2}\right\rangle }}{\left\langle a_{k}^{\dagger}a_{\mathrm{k}}a_{-k}^{\dagger}a_{\mathrm{-k}}\right\rangle }-1,
\end{equation}
as a measure. For coherent states, $I=0$, while for nonclassical
states we will have $I<0$ \cite{Agarwal1988,Agarwal2013}. 

In the Fig.$\,$2, we numerically investigate the nonclassical parameter
$I$ of the system in equilibrium state for various $n_{th}$ and
$G_{\mathrm{k}}$. We should notice that the blue-detuned drive produces
the photon-phonon pairs, and it would also induce instability of the
system due to the amplification. Using the well-known Routh-Hurwitz
stability conditions \cite{DeJesus1987}, we derive the necessary
and sufficient stability condition for this model as $G_{\mathrm{k}}^{2}-G_{-k}^{2}>-\kappa\gamma/4$.
In experiment, we can adjust the drive amplitude to guarantee $G_{\mathrm{-k}}^{2}<\kappa\gamma/4$,
therefore the system is always stable for arbitrary $G_{\mathrm{k}}$. 

The nonclassicality parameters $I$ as a function of the detuning
$\delta_{\mathrm{k}}=\omega_{s}-\omega_{c,k}$ for various thermal
phonon $n_{th}$ are plotted in Fig.$\,$2(b). As expected, the nonclassicality
parameter shows negativity $I<0$, which implies nonclassicality of
the two oppositely propagating fields due to the two drivings. When
the detuning $\delta_{\mathrm{k}}=0$, the nonclassicality parameter
reaches the smallest value. For more thermal phonon excitation $n_{th}$,
more noise phonon is converted to signal mode, thus the nonclassicality
reduces. This noise phonon conversion can also actually be suppressed
with the increasing of the effective coupling $G_{k}$, as shown in
Fig.$\,$2(c) that the nonclassicality is improved for larger $G_{k}$.
This is because when the effective coupling $G_{\mathrm{k}}$ increases,
the cooling of the phonon mode can be enhanced, and the thermal noise
is suppressed \cite{Aspelmeter2014}. As shown in Fig.$\,$2(d), we
plot the nonclassicality parameter $I$ as a function of $G_{\mathrm{k}}$
at the detuning $\delta_{k}=0$. The $I$ reduces with the coupling
$G_{\mathrm{k}}$ and approaches $-1$, which is the lower bound of
the parameter $I$. 

The degenerate CW and CCW signal photons travel in opposite directions,
thus the two mode can be separately coupled to different optical waveguides
{[}as shown in Fig.$\,$1(a){]}, which enable the type-II (spatially
separable) entanglement photon source on a photonic chip. Compared
to the crystals used in free space parametric down conversion, the
traveling-wave optomechanical approach has the advantages that narrow
linewidth, long coherence time, high purity and waveguide integrated.
This scheme can be used to realize quantum repeaters \cite{Simon2007}
and nonclassical interference \cite{Fulconis2007} by using indistinguishable
photon pair generated by the integrated microresonator. The potential
limitation of the thermal phonon $n_{th}$ can be relaxed by coupling
the mechanical mode to a third optical cavity mode, which is used
to cool the mechanical mode towards the ground state. When the damping
rates satify $\gamma_{m}\gg\kappa$, the steady-state entanglement
can be achieved via reservoir engineering \cite{Wang2013,Schmidt2012}.
For large mechanical noise $n_{th}\gg0$, quantum interference can
be applied to achieve robust photon entanglement \cite{Tian2013}
in the strong coupling regime $G>\kappa$.

\section{Phase-controlled non-reciprocity}

As shown in Fig.$\,$1(a), the signal photon propagating in the forward
and backward directions is coupling to the CCW and CW modes, therefore
directional pump to the system can induce transmission difference
between the forward and backward signal in the waveguide, which leads
to the optical non-reciprocity. Different from the single mechanical
mode induced non-reciprocity demonstrated in Ref. \cite{Shen2016,Ruesink2016},
here we propose a new scheme for phase-controlled non-reciprocal conversion
by two mechanical mode. The underlying mechanism is that the two mechanical
modes allow the paths for the coupling between the oppositely propagating
signals, and effective coupling can be controlled by changing the
relative phase of the two paths, which eventually leads to the one-way
conversion between two modes.The schematic of the scheme is shown
in Fig.$\,$3(a), where the CW and CCW modes are dispersively coupled
to two breath mechanical modes (one can be the radial expansion mode,
the other one can be the out-of-plain vibration mode). In the bosonic
four-mode system, the CW (CCW) optical mode is driven by lasers at
two frequencies $\omega_{c,k}-\omega_{m1}$ and $\omega_{c,k}-\omega_{m2}$,
with different amplitudes and phases. Under the condition that $\min\left[\omega_{m1},~\omega_{m2},~\left|\omega_{m1}-\omega_{m2}\right|\right]\gg$max$\left[\kappa,~\gamma_{m1},~\gamma_{m2}\right]$,
we can realize the coherent coupling between the phonons and photons,
the effective interaction Hamiltonian can be written as 
\begin{align}
H_{lin}= & \sum_{j=1,2}G_{kj}a_{k}^{\dagger}b_{0j}+G_{-kj}a_{-k}^{\dagger}b_{0j}+H.c.,
\end{align}
where $b_{0j}$ ($j=1,2$) is the annihilation operator of the mechanical
mode $j$, and $G_{kj}$ is the effective complex optomechanical coupling
with a certain phase from driving fields.

By including the damping effects, the dynamics of the system obeys
the quantum Langevin equations
\begin{equation}
\frac{dO}{dt}=-UO+\sqrt{\kappa}O_{in},
\end{equation}
where $O=\left(a_{k},~a_{-k},~b_{01},~b_{02}\right)^{T}$ is a vector
of the operators, the input vector $O_{in}=\left(\epsilon_{in,k},~\epsilon_{in,-k},~0,~0\right)^{T}$,
and the coefficient matrix
\begin{equation}
U=\left(\begin{array}{cccc}
\frac{\kappa}{2} & 0 & iG_{k1} & iG_{k2}\\
0 & \frac{\kappa}{2} & iG_{-k1} & iG_{-k2}\\
iG_{k1}^{\ast} & iG_{-k1}^{\ast} & \frac{\gamma_{m1}}{2} & 0\\
iG_{k2}^{\ast} & iG_{-k2}^{\ast} & 0 & \frac{\gamma_{m2}}{2}
\end{array}\right).
\end{equation}
Transforming the equations into the frequency domain, the intracavity
field is solved as $O=\left(U-i\omega I\right)^{-1}\sqrt{\kappa}O_{in}$.
Using the standard input-output theory $O_{out}=\sqrt{\kappa}O-O_{in}$,
the output of the system is obtained as $O_{out}=RO_{in}$, where
the scattering matrix $R=\kappa\left(U-i\omega I\right)^{-1}-I$.
We notice that there are two paths of the coherent conversion between
CW and CCW signal photons, therefore the total phase difference $\theta$
between the path $a_{k}\rightleftarrows b_{1}\rightleftarrows a_{-k}$
and the path $a_{k}\rightleftarrows b_{2}\rightleftarrows a_{-k}$
plays a key role in the optical mode conversion. When the symmetry
between the two paths is broken, the system becomes non-reciprocal.

\begin{figure}
\includegraphics[width=9cm]{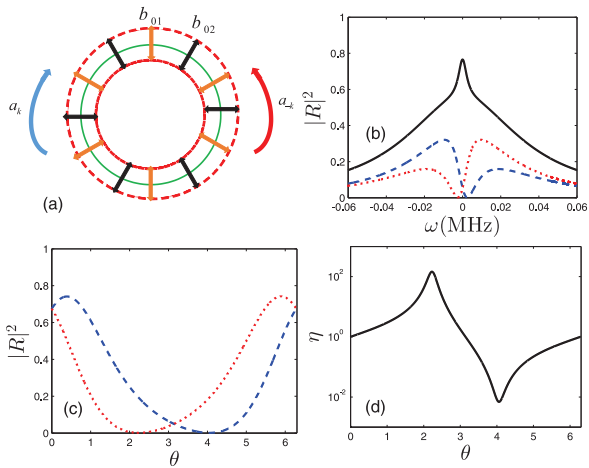}

\caption{(Color online) Optical mode conversion between CW and CCW fields.
(a) Schematic diagram of phase-controlled nonreciprocity. (b) The
conversion efficiency for the reciprocity $\theta=0$ (the black solid
line $\left|R_{k,-k}\right|^{2}=\left|R_{-k,k}\right|^{2}$) and nonreciprocity
$\theta=3\pi/4$ (the blue dashed line $\left|R_{k,-k}\right|^{2}$and
the red dotted line $\left|R_{-k,k}\right|^{2}$) as the function
of the signal frequency $\omega$. The conversional probabilities
(c) and conversional ratio $\eta$ (d) as the function of the phase
$\theta$ for the frequency $\omega=-\gamma_{m2}$. The other parameters
are $C_{k1}=C_{-k1}=1$, $C_{k2}=C_{-k2}=2.5$, $G_{k2}=\left|G_{k2}\right|e^{i\theta}$,
$\kappa/2\pi=15$ MHz, $\gamma_{m1}/2\pi=22\times10^{-3}$MHz, and
$\gamma_{m2}/2\pi=22\times10^{-4}$MHz.}
\end{figure}

To illustrate the phase-controlled non-reciprocity, we assume the
effective coupling strengths are real except for $G_{k2}=\left|G_{k2}\right|e^{i\theta}$.
For the convenience of expression, we define the cooperativity $C=\frac{4\left|G^{2}\right|}{\kappa\gamma}$.
If only one mechanical mode is coupled to the optical modes, it has
been demonstrated in the experiment that the transmission is non-reciprocal
only if $C_{k1}\neq C_{-k1}$, while the signal reflection is always
reciprocal \cite{Shen2016}. In additional to the optomechanically
induced phase shifting and absorption, the non-reciprocal conversion
between two modes is also very important for information processing,
and this can be realized in our scheme.

The Fig.$\,$3(b-d) shows the conversion efficiencies between CW and
CCW modes as a function of the signal frequency $\omega$ and the
pump phase differences $\theta$. The black solid line in the Fig.$\,$3(b)
shows the reciprocal conversion when the phase $\theta=0$. However,
when we choose the phase $\theta=3\pi/4$, as shown in Fig.$\,$3(b),
the nonreciprocal conversion $\left|R_{k,-k}\right|^{2}\neq\left|R_{-k,k}\right|^{2}$
is observed. When $\omega\approx-\gamma_{m2}$, we find $\left|R_{-k,k}\right|^{2}\rightarrow0$
(the red dotted line), which means that the conversion from the CW
to the CCW is almost forbidden, while the opposite conversion is feasible.
When $\omega\approx\gamma_{m2}$, there is an opposite result, which
is confirmed by the blue dashed line $\left|R_{k,-k}\right|^{2}$.
To further illustrate the effect of the phase, we plot the conversion
efficiency as the function of the phase $\theta$ for the frequency
$\omega=-\gamma_{m2}$ in the Fig.$\,$3(c), which obviously shows
that the optical conversion can be periodically changed with the increasing
of the phase $\theta$. Fig.$\,$3(d) shows the ratio $\eta=\left|R_{k,-k}\right|^{2}/\left|R_{-k,k}\right|^{2}$as
the function of the phase $\theta$, which can reach the nonreciprocal
degree $\eta\propto10^{2}$ for the experimental parameters. It is
obvious that the total phase difference $\theta$ plays a key role
in the optical mode conversion.

If we can realize the parameters $\gamma_{m2}\ll G\sim\kappa\ll\gamma_{m1}$,
the ideal non-reciprocal optical conversion is possible. This assumption
seems to be counter-intuitive because usually the damping rate of
the mechanical mode is smaller than the decay rate of the cavity mode,
but we can satisfy the condition when the mechanical mode is coupled
to an auxiliary cavity mode \cite{Xu2016}. In addition, the signal
frequency $\omega$ of ideal optical conversion can be tuned by the
detuning between the driving fields and the optical modes, which has
been recently demonstrated in the microwave transmission \cite{Bernier2016}.

\section{Phononic parity-time symmetry}

\begin{figure}[H]
\includegraphics[width=9cm]{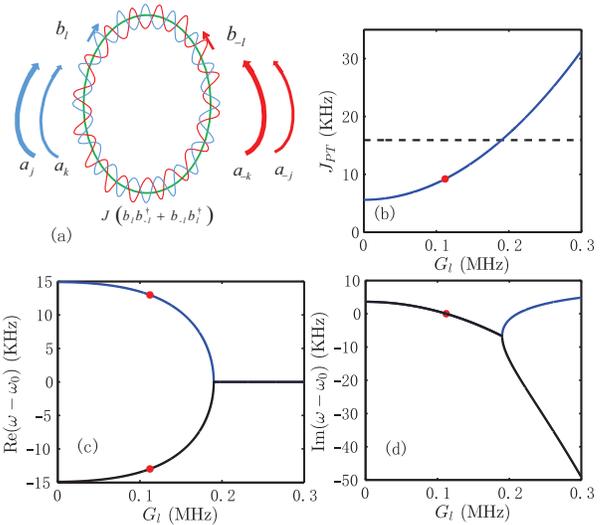}

\caption{(Color online) PT-symmetry in deformation optomechanical system. (a)
Schematic diagram of PT-symmetry. The threshold coupling strength
$J_{PT}$ (b), the real Re$\left(\omega\right)$ (c) and Im$\left(\omega\right)$
(d) versus the $G_{l}$. The black dash line in (b) shows the coupling
$J/2\pi=16$ KHz. The red dots means that gain and loss are balanced
(i.e., $\gamma_{-l}+\gamma_{m}=-\gamma_{l}-\gamma_{\mathrm{m}}$),
which is the ideal PT-symmetry. The parameters are $\omega_{ml}/2\pi=42.3$
MHz, $\gamma_{\mathrm{\mathrm{m}}}/2\pi=4$ KHz, $G_{-l}/2\pi=0.14$
MHz, $J/2\pi=16$ KHz, and $\kappa_{\mathrm{\mathrm{1}}}/2\pi=\kappa_{\mathrm{2}}/2\pi=3.5$
MHz.}
\end{figure}
Most of previous studies are focus on the control of multiple optical
signal modes. The degenerate CW and CCW acoustic modes can also be
used for interesting applications. Here, we consider a stimulated
Brillouin scattering in this triple-resonant optomechanics \cite{Kim2015,Dong2015},
in which the Hamiltonian can be described as 
\begin{flalign}
H_{sbs} & =\sum_{o=j,k}\omega_{c,o}a_{o}^{\dagger}a_{o}+\omega_{ml}b_{l}^{\dagger}b_{l}+g_{j,k,l}\left(b_{l}^{\dagger}a_{\mathrm{j}}^{\dagger}a_{k}+H.c\right),
\end{flalign}
where $a_{j}$, $a_{k}$ are the optical modes and $b_{l}$ is the
acoustic mode that satisfies the selection rule. 

In a perfect rotational symmetric whispering-gallery microresonator,
the CW and CCW acoustic modes are degenerate. To study the non-Hermitian
physics of the phononic mode, we consider a slightly deformed microresonator
that breaks the rotational symmetry and induces an effective coupling
between $b_{l}$ and $b_{-l}$. The schematic of the system is shown
in the Fig.$\,$4(a), and the phonon coupling is described as
\begin{equation}
H_{bs}=J(b_{l}^{\dagger}b_{-l}+b_{l}b_{-l}^{\dagger}),
\end{equation}
where $J$ is the deformed boundary induced backscattering, i.e. the
coupling between the mechanical CW and CCW modes. In practical experimental
system, the $J$ can be easily controlled to be the order of $\gamma_{m}$
without degrading the high quality factor of optical and mechanical
modes. Since the optical wavelength is much shorter than that of acoustic
modes, the smooth boundary deformation will not induce observable
backscattering between the optical modes.

If we drive the optical mode $a_{j}$ for the CW direction and $a_{-k}$
for the CCW direction by two control fields, the Hamiltonian can be
simplified as
\begin{align}
H_{lin}= & H_{0}+J(b_{l}^{\dagger}b_{-l}+b_{l}b_{-l}^{\dagger})\nonumber \\
 & +\left(G_{l}a_{k}^{\dagger}b_{l}+G_{-l}a_{-j}b_{-l}+H.c\right),
\end{align}
where $G_{l\left(-l\right)}=g_{j,k,l}\alpha_{j\left(-k\right)}$ is
the effective optomechanical coupling by the driving field $\alpha_{j\left(-k\right)}$.The
Hamiltonian $H_{0}=\Delta_{k}a_{k}^{\dagger}a_{k}+\Delta_{-j}a_{-j}^{\dagger}a_{-j}+\omega_{ml}\left(b_{l}^{\dagger}b_{l}+b_{-l}^{\dagger}b_{-l}\right)$,
in which $\Delta_{k}=\omega_{c,k}-\omega_{j,l}$ and $\Delta_{-j}=\omega_{c,j}-\omega_{\mathrm{-k},l}$,
where $\omega_{j,l}$ and $\omega_{-k,l}$ are the frequencies driving
on $a_{j}$ and $a_{-k}$, respectively. 

To investigate the PT-symmetric mechnaical system \cite{Jing2014,Konotop2016},
we use a probe field to drive the optical mode $a_{-j}$ with the
frequency $\omega_{p}$. We can write the following rate equations
for the coupled-resonators system
\begin{equation}
\frac{dO}{dt}=-UO+\sqrt{\kappa_{in}}O_{in},\label{Eq:PTLangivin}
\end{equation}
where $O=\left(a_{k},~a_{-j}^{\dagger},~b_{l},~b_{-l}\right)^{T}$
is a vector of the operators, the input vector $O_{in}=\left(0,~\epsilon_{p},~0,~0\right)^{T}$,
and the coefficient matrix
\begin{equation}
U=\left(\begin{array}{cccc}
i\delta_{k}+\frac{\kappa_{2}}{2} & 0 & iG_{l} & 0\\
0 & -i\delta_{-j}+\frac{\kappa_{1}}{2} & 0 & -iG_{-l}\\
iG_{l} & 0 & i\delta_{l}+\frac{\gamma_{m}}{2} & iJ\\
0 & iG_{-l} & iJ & i\delta_{-l}+\frac{\gamma_{m}}{2}
\end{array}\right),
\end{equation}
where $\delta_{k}=\Delta_{k}+\omega_{p}-\omega_{-k,l}$, $\delta_{-j}=\omega_{c,j}-\omega_{p}$,
$\delta_{l}=\omega_{ml}+\omega_{p}-\omega_{-k,l}$, $\delta_{-l}=\delta_{l}$.
For $\kappa_{1,2}\gg\gamma_{\mathrm{\mathrm{m}}}$, we can adiabatically
eliminate the optical modes by
\begin{align}
a_{-j}^{\dagger} & =\frac{iG_{-l}b_{-l}+\sqrt{\kappa_{in}}\epsilon_{p}}{-i\delta_{-j}+\frac{\kappa_{1}}{2}},\\
a_{k} & =\frac{-iG_{l}b_{l}}{i\delta_{k}+\frac{\kappa_{2}}{2}},
\end{align}
and we can obtain the effective rate equations of the acoustic modes
\begin{eqnarray}
\frac{db_{l}}{dt} & = & -\left(i\delta_{l}+\frac{\gamma_{m}}{2}+\frac{G_{l}^{2}}{i\delta_{k}+\frac{\kappa_{2}}{2}}\right)b_{l}\nonumber \\
 &  & -iJb_{-l},\\
\frac{db_{-l}}{dt} & = & -\left(i\delta_{-l}+\frac{\gamma_{m}}{2}-\frac{G_{-l}^{2}}{-i\delta_{-j}+\frac{\kappa_{1}}{2}}\right)b_{-l}\nonumber \\
 &  & -iJb_{l}-\frac{iG_{-l}\sqrt{\kappa_{in}}\epsilon_{p}}{-i\delta_{-j}+\frac{\kappa_{1}}{2}}.
\end{eqnarray}
Here we choose the driving frequency $\omega_{-k,l}=\omega_{c,k}$
and $\omega_{j,l}=\omega_{c,j}=\omega_{p}$. The coupling of these
two resonators creates two supermodes $B_{+}=\left(b_{l}+b_{-l}\right)/\sqrt{2}$
and $B_{-}=\left(b_{-l}-b_{l}\right)/\sqrt{2}$ with the eigenfrequencies
$\omega_{+}$ and $\omega_{-}$ given as 
\begin{equation}
\omega_{\pm}=\omega_{0}-\frac{i}{4}\left(2\gamma_{m}+\gamma_{-l}+\gamma_{l}\right)\pm\frac{1}{4}\sqrt{16J^{2}-\left(\gamma_{-l}-\gamma_{l}\right)^{2}}.
\end{equation}
where $\gamma_{-l}=-\frac{4G_{-l}^{2}}{\kappa_{1}}$, $\gamma_{l}=\frac{4G_{l}^{2}}{\kappa_{2}}$,
and $\omega_{0}=\frac{\delta_{l}+\delta_{-l}}{2}$.

Then the value of $J$ sastisfying $16J^{2}=\left(\gamma_{-l}-\gamma_{l}\right)^{2}$
is the threshold coupling strength $J_{PT}$, which is found as
\begin{equation}
J_{PT}=\frac{1}{4}\left|\gamma_{-l}-\gamma_{l}\right|.
\end{equation}
It is clear that $J_{PT}$ depends on the drving strength of the system.
For the ideal case when gain and loss are balanced (i.e., $\gamma_{-l}+\gamma_{\mathrm{\mathrm{m}}}=-\gamma_{l}-\gamma_{\mathrm{m}}$),
the coupling strength $J_{PT}$ becomes 
\begin{equation}
J_{PT}=\frac{\gamma_{l}+\gamma_{\mathrm{m}}}{2}.
\end{equation}

\begin{figure}[H]
\includegraphics[width=9cm]{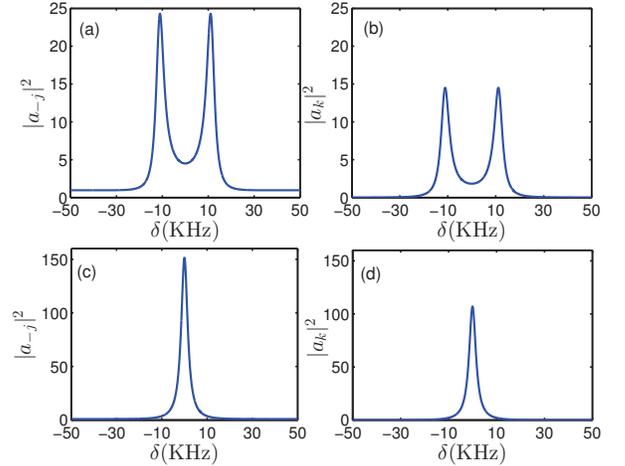}

\caption{The spectra as the function of the detuning $\delta$. (a) $\left|a_{-j}\right|^{2}$
and (b) $\left|a_{k}\right|^{2}$ with $G_{l}/2\pi=0.14$ MHz. We
have $J_{PT}<J$ , which is related to the unbroken-PT-symmetry. (c)
$\left|a_{-j}\right|^{2}$ and (d) $\left|a_{k}\right|^{2}$ show
the broken-PT-symmetry for $J_{PT}>J$ with $G_{l}/2\pi=0.2$ MHz.
Other parameters are the same as the above.}
\end{figure}
The threshold coupling strength $J_{PT}$ versus the $G_{l}$ for
the fixed $G_{-l}/2\pi=0.14$ MHz is plotted in the Fig.$\,$4(b).
The black dash line shows the coupling $J/2\pi=16\,\mathrm{kHz}$
and the red dot is corresponding to the ideal PT-symmetry. For $J_{PT}/J<1$,
the system is in a unbroken-symmetry phase in Fig.$\,$4(c) and 4(d),
as seen in both the non-zero difference of the real parts of the eigenfrequencies
and the coalescence in their imaginary parts. As $J_{PT}/J$ approaches
1 from below, the difference in the real parts of the eigenfrequencies
decreases and their imaginary parts bifurcate.

To observe the phenomena of the PT-symmetry, we can calculate the
intracavity field by the Eq.$\,$(\ref{Eq:PTLangivin}). When the
system is in the steady state $\frac{dO}{dt}=0$, we obtain 
\begin{align}
a_{k} & =\frac{-iG_{l}b_{l}}{i\delta+\kappa_{2}/2},\\
a_{-j}^{\dagger} & =\frac{iG_{-l}b_{-l}+\sqrt{\kappa_{in}}\epsilon_{p}}{i\delta+\frac{\kappa_{1}}{2}},\\
b_{l} & =\frac{iJb_{-l}}{F_{2}\left(\delta\right)},\\
b_{-l} & =\frac{iG_{-l}\sqrt{\kappa_{ex}}\epsilon_{p}}{i\delta+\frac{\kappa_{1}}{2}}/\left[F_{1}\left(\delta\right)+\frac{J^{2}}{F_{2}\left(\delta\right)}\right],
\end{align}
where $F_{1}\left(\delta\right)=-i\delta-\frac{\gamma_{m}}{2}+\frac{G_{-l}^{2}}{i\delta+\frac{\kappa_{1}}{2}}$
and $F_{2}\left(\delta\right)=-i\delta-\frac{\gamma_{\mathrm{m}}}{2}-\frac{G_{l}^{2}}{i\delta+\kappa_{2}/2}$,
in which the detuning $\delta=\omega_{p}-\omega_{c,j}$.

The spectra of the steady-state intracavity field are shown in Fig.$\,$5.
When we choose the coupling $G_{l}/2\pi=0.14$ MHz, we have the threshold
coupling strength $J_{PT}<J$ , which is in the unbroken-PT-symmetric
region. Fig.$\,$5(a) shows the intracavity field $\left|a_{-j}\right|^{2}$
as the function of detuning $\delta$, and there are two resonance
peaks, which are corresponding to non-zero difference of the real
parts of the eigenfrequencies in the Fig.$\,$4(b). The imaginary
parts of the eigenfrequencies in Fig. 4(c) show the coalescence, which
is confirmed with the same width of the two peaks. The intracavity
field $\left|a_{k}\right|^{2}$ is plotted in the Fig.$\,$5(b), which
shows the similar phenomena. When $J_{PT}>J$ with the coupling $G_{l}/2\pi=0.2$
MHz, we observe the broken-PT-symmetric phenomena, and there is one
peak in the Fig.$\,$5(c) and 5(d). By the deformation optomechanical
system, we demonstrate that the parity-time symmetry of acoustic modes
can be observed. More importantly, gain could be provided only by
optical fields without nonlinear processes. Based on parameters realized
with existing technology in recent experiments, these optomechanical
devices should be readily achievable. 

The parameters we choose above satisfy Im$\left(\omega-\omega_{0}\right)<0$,
which means that all modes have the loss. When Im$\left(\omega-\omega_{0}\right)>0$,
it is obvious that the system is unstable, and we can calculate the
intracavity field with driving pulse numerically by the Eq.$\,$(\ref{Eq:PTLangivin}).
In the linear regime, the transmission is reciprocal regardless of
whether the symmetry is broken or unbroken.

\section{Discussion}

The traveling-wave optomechanical system can be realized in the solid
state whispering gallery microresonators, with various shapes and
materials \cite{Strekalov2016}. Those high-quality materials usually
support not only high quality factor ($Q$) optical modes, but also
supports high $Q$ mechanical modes. Especially, in polished single
crystal resonators, the mechanical quality factor can be as high as
$10^{5}$ \cite{Zhang2016}. Therefore, this type microresonator has
great potential for the coherent optomechanical interactions. 

Compared with the optical modes, the mechanical modes have much longer
lifetime and can be used as quantum memory. In compare with the usual
single mode optomechanics, the traveling-wave microresoantors have
two unique advantages, the mode degeneracy and suppression of dephasing.
By employing the degenerate CW and CCW acoustic modes, the binary
bosonic memory is possible \cite{Dong2015}. The degenerate traveling
modes have the remarkable properties that the modes bear the same
noise source, and the noises cancel in most of applications. For example,
if a single phonon excitation is stored as a superposition of CW and
CCW directions, their phase coherence can be insensitive to the external
vibration and thermal noise. Because the dominated dephasing noise
of both CW and CCW are the same, the relative phase between CW and
CCW mode conserves. Therefore, the natural degenerate modes in traveling-wave
microresonators form a decoherence-free subspace and preserve the
coherence.

In practical applications, the degeneracy of the CW and CCW modes
would also save the laser source. We can use one pump laser to study
the coherent interaction in two directions. Besides the CW and CCW
modes studied in this work, the traveling-wave optomechanical system
supports optical modes and mechanical modes in a large frequency range,
with the mode frequency approximately proportional to the angular
momentum $\left|l\right|$. By choosing a proper material, there will
be hundreds of optical modes covering the visible, IR and mid-IR frequency
range. Similarly, the acoustic wave modes also show a comb like spectrum,
and have the frequency range from few MHz to GHz \cite{Savchenkov2011}.
So, the applications discussed in this paper on the CW and CCW modes
can be generalized to the modes with very different frequencies.

\section{Conclusion}

In summary, we theoretically study the multimode optomechanics based
on the traveling-wave microresonators, where the degenerate clockwise
and counter-clockwise optical and acoustical modes can be controlled
selectively by optically pumping in different directions. By the unique
properties of the traveling-wave optomechanic systems, we have proposed
several unique applications of the system. Based on the parameters
realized in recent experiments, these optomechanical devices should
be readily achievable, even on the level of single-photon with practical
noises.

{\em Acknowledgments.} We thank Yuan Chen for preparing figure
1b and 1c, and acknowledge Chuan-Shen Yang and Hui Jing for useful
suggestions. This work was funded by the National Key R \& D Program
(Grant No. 2016YFA0301300, and 2016YFA0301700), and the National Natural
Science Foundation of China (Grant No. 11474271 and Grant No. 11674305),
the China Postdoctoral Science Foundation (No. 2016M602013). YDW acknowledges
the support of NSFC grant No. 11574330 and No. 11434011.


\begin{thebibliography}{10}
\bibitem{Aspelmeter2014} M. Aspelmeyer, T. J. Kippenberg, and F.
Marquardt, Cavity optomechanics, Rev. Mod. Phys. \textbf{86}, 1391
(2014).

\bibitem{Kippenberg2008} T. J. Kippenberg and K. J. Vahala, Cavity
Optomechanics: Back-Action at the Mesoscale, Science \textbf{321},
1172 (2008).

\bibitem{Aspelmeyer2012} M. Aspelmeyer, P. Meystre, and K. Schwab,
Quantum optomechanics, Phys. Today \textbf{65}, 29 (2012).

\bibitem{Thourhout2010} D. Van Thourhout and J. Roels, Optomechanical
device actuation through the optical gradient force, Nat. Photon.
\textbf{4}, 211 (2010).

\bibitem{Hong2013} T. Hong, H. Yang, H. Miao, and Y. Chen, Open quantum
dynamics of single-photon optomechanical devices, Phys. Rev. A \textbf{88},
023812 (2013).

\bibitem{Tang1} M. Li, W. H. P. Pernice, C. Xiong, T. Baehr-Jones,
M. Hochberg, and H. X. Tang, Harnessing optical forces in integrated
photonic circuits, Nature \textbf{456}, 480 (2008).

\bibitem{Eichenfield2009} M. Eichenfield, R. Camacho, J. Chan, K.
J. Vahala, and O. Painter, A picogram- and nanometre-scale photonic-crystal
optomechanical cavity, Nature \textbf{459}, 550 (2009).

\bibitem{Fan2016} L. Fan, C.-L. Zou, M. Poot, R. Cheng, X. Guo, X.
Han, and H. X. Tang, Integrated optomechanical single-photon frequency
shifter, Nat. Photonics \textbf{10}, 766\textendash 770 (2016).

\bibitem{Tang2} M. Bagheri, M. Poot, M. Li, W. P. H. Pernice, and
H. X. Tang, Dynamic manipulation of nanomechanical resonators in the
high-amplitude regime and non-volatile mechanical memory operation,
Nat. Nanotechnol. \textbf{6}, 726 (2011).

\bibitem{Cole2011} G. D. Cole, and M. Aspelmeyer, Cavity optomechanics:
Mechanical memory sees the light, Nat. Nanotechnol. \textbf{6}, 690
(2011).

\bibitem{Anetsberger2010} G. Anetsberger, E. Gavartin, O. Arcizet,
Q. P. Unterreithmeier, E. M. Weig, M. L. Gorodetsky, J. P. Kotthaus,
and T. J. Kippenberg, Measuring nanomechanical motion with an imprecision
below the standard quantum limit, Phys. Rev. A \textbf{82}, 061804
(2010).

\bibitem{Weber2016} P. Weber, J. Güttinger, A. Noury, J. Vergara-Cruz
, and A. Bachtold, Force sensitivity of multilayer graphene optomechanical
devices, Nat. Commun. \textbf{7}, 12496 (2016).

\bibitem{Ockeloen-Korppi2016} C.\LyXThinSpace F. Ockeloen-Korppi,
E. Damskägg, J.-M. Pirkkalainen, A.\LyXThinSpace A. Clerk, M.\LyXThinSpace J.
Woolley, and M.\LyXThinSpace A. Sillanpää, Quantum Backaction Evading
Measurement of Collective Mechanical Modes, Phys. Rev. Lett. \textbf{117},
140401 (2016).

\bibitem{Aasi2013} J. Aasi et al., Enhanced sensitivity of the LIGO
gravitational wave detector by using squeezed states of light, Nat.
Photon. \textbf{7}, 613 (2013).

\bibitem{Belenchia2016} A. Belenchia, D. M.\LyXThinSpace T. Benincasa,
S. Liberati, F. Marin, F. Marino, and A. Ortolan, Testing Quantum
Gravity Induced Nonlocality via Optomechanical Quantum Oscillators,
Phys. Rev. Lett. \textbf{116}, 161303 (2016).

\bibitem{Ludwig2008} M. Ludwig, B. Kubala, and F. Marquardt, The
optomechanical instability in the quantum regime, New J. Phys. \textbf{10},
095013 (2008).

\bibitem{Brennecke2008} F. Brennecke, S. Ritter, T. Donner, T. Esslinger,
Cavity Optomechanics with a Bose-Einstein Condensate, Science \textbf{322},
235 (2008).

\bibitem{Khalili2010} F. Khalili, S. Danilishin, H. Miao, H. Müller-Ebhardt,
H. Yang, and Y. Chen, Preparing a Mechanical Oscillator in Non-Gaussian
Quantum States, Phys. Rev. Lett. \textbf{105}, 070403 (2010).

\bibitem{Stannigel2012} K. Stannigel, P. Komar, S. J. M. Habraken,
S. D. Bennett, M. D. Lukin, P. Zoller, and P. Rabl, Optomechanical
Quantum Information Processing with Photons and Phonons, Phys. Rev.
Lett. \textbf{109}, 013603 (2012).

\bibitem{Purdy2013} T. P. Purdy, R. W. Peterson, and C. A. Regal,
Observation of Radiation Pressure Shot Noise on a Macroscopic Object,
Science \textbf{339}, 801(2013).

\bibitem{Teufel2011} J. D. Teufel et al., Sideband cooling of micromechanical
motion to the quantum ground state, Nature \textbf{475}, 359 (2011).

\bibitem{Weis2010} S. Weis, R. Riviére, S. Deléglise, E. Gavartin,
O. Arcizet, A. Schliesser, and T. J. Kippenberg, Optomechanically
Induced Transparency, Science \textbf{330}, 1520 (2010).

\bibitem{Safavi-Naeini2011} A. H. Safavi-Naeini, T. P. Mayer Alegre,
J. Chan, M. Eichenfield, M. Winger, Q. Lin, J. T. Hill, D. E. Chang,
and O. Painter, Electromagnetically induced transparency and slow
light with optomechanics, Nature \textbf{472,} 69 (2011).

\bibitem{Palomaki2013} T.\LyXThinSpace A. Palomaki, J.\LyXThinSpace D.
Teufel, R.\LyXThinSpace W. Simmonds, and K.\LyXThinSpace W. Lehnert,
Entangling Mechanical Motion with Microwave Fields, Science \textbf{342},
710 (2013).

\bibitem{Wollman2015} E.\LyXThinSpace E. Wollman, C.\LyXThinSpace U.
Lei, A.\LyXThinSpace J. Weinstein, J. Suh, A. Kronwald, F. Marquart,
A.\LyXThinSpace A. Clerk, and K.\LyXThinSpace C. Schwab, Quantum squeezing
of motion in a mechanical resonator, Science \textbf{349}, 952 (2015).

\bibitem{Dong2012} C. Dong, V. Fiore, M.\LyXThinSpace C. Kuzyk, and
H. Wang, Optomechanical Dark Mode, Science \textbf{338}, 1609 (2012).

\bibitem{Andrews2014} R. W. Andrews, R. W. Peterson, T. P. Purdy,
K. Cicak, R. W. Simmonds, C. A. Regal, and K. W. Lehnert, Bidirectional
and efficient conversion between microwave and optical light, Nat.
Phys. \textbf{10}, 321 (2014).

\bibitem{Xu2015} X.-W. Xu and Y. Li, Optical nonreciprocity and optomechanical
circulator in three-mode optomechanical systems, Phys. Rev. A \textbf{91},
053854 (2015).

\bibitem{Meltelmann2015} A. Meltelmann and A. A. Clerk, Nonreciprocal
Photon Transmission and Amplification via Reservoir Engineering, Phys.
Rev. X \textbf{5}, 021025 (2015).

\bibitem{Peano2016} V. Peano, M. Houde, C. Brendel, F. Marquardt,
and A. A. Clerk, Topological phase transitions and chiral inelastic
transport induced by the squeezing of light, Nat. Commun. \textbf{7},
10779 (2016).

\bibitem{Peng2014}B. Peng et al., Parity\textendash time-symmetric
whispering-gallery microcavities, Nat. Phys. \textbf{10}, 394 (2014).

\bibitem{Wurl2016} C. Wurl, A. Alvermann, and H. Fehske, Symmetry-breaking
oscillations in membrane optomechanics, Phys. Rev. A \textbf{94} 063860
(2016).

\bibitem{HXu2016} H. Xu, D. Mason, Luyao Jiang, and J. G. E. Harris,
Topological energy transfer in an optomechanical system with exceptional
points, Nature \textbf{537}, 80 (2016).

\bibitem{phononlaser} I. S. Grudinin, H. Lee, O. Painter, and K.
J. Vahala, Phonon Laser Action in a Tunable Two-Level System, Phys.
Rev. Lett. \textbf{104}, 083901 (2010).

\bibitem{Zhu} J. Fan and L. Zhu, Enhanced optomechanical interaction
in coupled microresonators, Opt. Express \textbf{20}, 20790 (2012).

\bibitem{Buchmann2015} L. F. Buchmann and D. M. Stamper-Kurn, Nondegenerate
multimode optomechanics, Phys. Rev. A \textbf{92}, 013851 (2015).

\bibitem{Kipf2014} T. Kipf and G. S. Agarwal, Superradiance and collective
gain in multimode optomechanics, Phys. Rev. A \textbf{90}, 053808
(2014).

\bibitem{Chesi2015} S. Chesi, Y. D. Wang, and J. Twamley, Diabolical
points in multi-scatterer optomechanical systems, Sci. Rep. \textbf{5},
7816 (2015).

\bibitem{Deng2016} Z. J. Deng, X.-B. Yan, Y.-D. Wang, and C.-W. Wu,
Optimizing the output-photon entanglement in multimode optomechanical
systems, Phys. Rev. A \textbf{93}, 033842 (2016).

\bibitem{Park2009} Y.-S. Park and H. Wang, Resolved-sideband and
cryogenic cooling of an optomechanical resonator, Nat. Phys. \textbf{5},
489 (2009).

\bibitem{Balram2014} K. C. Balram, M. Davanço, J. Y. Lim, J. D. Song,
and K. Srinivasan, Moving boundary and photoelastic coupling in GaAs
optomechanical resonators, Optica \textbf{1}, 414 (2014).

\bibitem{Baker2014} C. Baker, W. Hease, D.-T. Nguyen, A. Andronico,
S. Ducci, G. Leo, and I. Favero, Photoelastic coupling in gallium
arsenide optomechanical disk resonators, Opt. Express \textbf{22},
14072 (2014).

\bibitem{Tomes2011} M. Tomes, F. Marquardt, G. Bahl, and T. Carmon,
Quantum-mechanical theory of optomechanical Brillouin cooling, Phys.
Rev. A \textbf{84}, 063806 (2011). 

\bibitem{Grudinin2009} I. S. Grudinin, A. B. Matsko, and L. Maleki,
Brillouin Lasing with a CaF2 Whispering Gallery Mode Resonator, Phys.
Rev. Lett. \textbf{102}, 043902 (2009).

\bibitem{Bahl2011} G. Bahl, J. Zehnpfennig, M. Tomes, and T. Carmon,
Stimulated optomechanical excitation of surface acoustic waves in
a microdevice. Nat. Commun. \textbf{2}, 403 (2011).

\bibitem{Bahl2012} G. Bahl, M. Tomes, F. Marquardt, and T. Carmon,
Observation of spontaneous Brillouin cooling. Nat. Phys. \textbf{8},
203\textendash 207 (2012).

\bibitem{Shen2016} Z. Shen, Y.-L. Zhang, Y. Chen, C.-L. Zou, Y.-F.
Xiao, X.-B. Zou, F.-W. Sun, G.-C. Guo, and C.-H. Dong, Experimental
realization of optomechanically induced non-reciprocity, Nat. Photon.
\textbf{10}, 657 (2016).

\bibitem{Ruesink2016} F. Ruesink, M.-A. Miri, A. Alù, and E. Verhagen,
Nonreciprocity and magnetic-free isolation based on optomechanical
interactions, Nat. Commun. \textbf{7}, 13662 (2016).

\bibitem{Kim2015} J. H. Kim, M C. Kuzyk, K. Han, H. Wang, and G.
Bahl, Non-reciprocal Brillouin scattering induced transparency, Nat.
Phys. \textbf{11}, 275 (2015).

\bibitem{Dong2015} C.-H. Dong, Z. Shen, C.-L. Zou, Y.-L. Zhang, W.
Fu, and G.-C. Guo, Brillouin-scattering-induced transparency and non-reciprocal
light storage, Nat. Commun. \textbf{6}, 6193 (2015).

\bibitem{Scully1997} M. O. Scully and M. S. Zubairy, Quantum Optics
(Cambridge Univ. Press, Cambridge, 1997).

\bibitem{Agarwal1988} G. S. Agarwal, Nonclassical statistics of fields
in pair coherent states, J. Opt. Soc. Am. B \textbf{5}, 1940 (1988).

\bibitem{Agarwal2013} Girish S. Agarwal, Quantum Optics (Cambridge
Univ. Press, Cambridge, 2013).

\bibitem{DeJesus1987} E. X. DeJesus and C. Kaufman, Routh-Hurwitz
criterion in the examination of eigenvalues of a system of nonlinear
ordinary differential equations, Phys. Rev. A \textbf{35}, 5288 (1987).

\bibitem{Simon2007} C. Simon, H. de Riedmatten, M. Afzelius, N. Sangouard,
H. Zbinden, and N. Gisin, Quantum Repeaters with Photon Pair Sources
and Multimode Memories, Phys. Rev. Lett. \textbf{98}, 190503 (2007).

\bibitem{Fulconis2007} J. Fulconis, O. Alibart, J. L. O\textquoteright Brien,
W. J. Wadsworth, and J. G. Rarity, Nonclassical Interference and Entanglement
Generation Using a Photonic Crystal Fiber Pair Photon Source, Phys.
Rev. Lett. \textbf{99}, 120501 (2007).

\bibitem{Wang2013} Ying-Dan Wang and Aashish A. Clerk, Reservoir-Engineered
Entanglement in Optomechanical Systems, Phys. Rev. Lett. \textbf{110},
253601 (2013).

\bibitem{Schmidt2012}M. Schmidt, M. Ludwig, and F. Marquardt, Optomechanical
circuits for nanomechanical continuous variable quantum state processing,
New J. Phys. \textbf{14}, 125005 (2012).

\bibitem{Tian2013} Lin Tian, Robust Photon Entanglement via Quantum
Interference in Optomechanical Interfaces, Phys. Rev. Lett. \textbf{110},
233602 (2013).

\bibitem{Xu2016}Xun-Wei Xu, Yong Li, Ai-Xi Chen, and Yu-xi Liu, Nonreciprocal
conversion between microwave and optical photons in electro-optomechanical
systems, Phys. Rev. A \textbf{93}, 023827 (2016).

\bibitem{Bernier2016} N. R. Bernier, L. D. Tóth, A. Koottandavida,
A. Nunnenkamp, A. K. Feofanov, T. J. Kippenberg, Nonreciprocal reconfigurable
microwave optomechanical circuit, arXiv:1612.08223 (2016).

\bibitem{Jing2014} H. Jing, S.\LyXThinSpace K. Özdemir, X.-Y. Lü,
J. Zhang, L. Yang, and F. Nori, PT -Symmetric Phonon Lase, Phys. Rev.
Lett. \textbf{113}, 053604 (2014).

\bibitem{Konotop2016} V. V. Konotop, J. Yang, and D. A. Zezyulin,
Nonlinear waves in PT -symmetric systems, Rev. Mod. Phys. \textbf{88},
035002 (2016).

\bibitem{Strekalov2016} D. V Strekalov, C. Marquardt, A. B. Matsko,
H. G. L. Schwefel, and G. Leuchs, Nonlinear and quantum optics with
whispering gallery resonators, J. Opt. \textbf{18}, 123002 (2016).

\bibitem{Zhang2016} X. Zhang, C.-L. Zou, L. Jiang, and H. X. Tang,
Cavity magnomechanics, Science Advances \textbf{2}, e1501286 (2016).

\bibitem{Savchenkov2011} A. A. Savchenkov, A. B. Matsko, V. S. Ilchenko,
D. Seidel, and L. Maleki, Surface acoustic wave opto-mechanical oscillator
and frequency comb generator, Opt. Lett. \textbf{36}, 3338\textendash 40
(2011).
\end{thebibliography}
\end{document}